# Toward the Automatic Generation of a Semantic VRML Model from Unorganized 3D Point Clouds

Helmi Ben Hmida, Frank Boochs
Institut i3mainz, am Fachbereich Geoinformatik und Vermessung
Fachhochschule Mainz, Lucy-Hillebrand-Str. 255128 Mainz, Germany
e-mail: {helmi.benhmida, boochs}@geoinform.fh-mainz.de

Christophe Cruz, Christophe Nicolle
Laboratoire Le2i, UFR Sciences et Techniques

Université de Bourgogne
B.P. 47870, 21078 Dijon Cedex, France
e-mail: {christophe.cruz, cnicolle}@u-bourgogne.fr

*Abstract*—This paper presents our experience regarding the creation of 3D semantic facility model out of unorganized 3D point clouds. Thus, a knowledge-based detection approach of objects using the OWL ontology language is presented. This knowledge is used to define SWRL detection rules. In addition, the combination of 3D processing built-ins and topological Built-Ins in SWRL rules aims at combining geometrical analysis of 3D point clouds and specialist's knowledge. This combination allows more flexible and intelligent detection and the annotation of objects contained in 3D point clouds. The created WiDOP prototype takes a set of 3D point clouds as input, and produces an indexed scene of colored objects visualized within VRML language as output. The context of the study is the detection of railway objects materialized within the Deutsche Bahn scene such as signals, technical cupboards, electric poles, etc. Therefore, the resulting enriched and populated domain ontology, that contains the annotations of objects in the point clouds, is used to feed a GIS system.

## I. INTRODUCTION

The technical survey of facility aims to build a digital model based on geometric analysis. Such a process becomes more and more tedious, especially with the generation of the new terrestrial laser scanners, faster, accurate, where huge amount of 3D point clouds is generated. Within such new technologies, new challenges have seen the light where the basic one is to make the reconstruction process automatic and more accurate. Thus, early works on 3D point clouds have investigated the reconstruction and the recognition of geometrical shapes [*1*], [*2*]. This problematic was investigated as a topic of the computer graphic and the signal processing research where most works focused on segmentation or visualization aspects. As most recent works, new tendency related to the use of semantic has been explored [*3*]. In fact, we agree with the assumption that it helps the improvement of the automation, the accuracy and the result quality, but we think that it has to be well studied and proved. Otherwise, how the detection process can get support within different knowledge about the scene objects and what´s its impact compared to classic approach. In such scenario, knowledge about such objects has to include detailed information about the objects geometry, structure, 3D algorithms, etc.

By this contribution, we suggest a proposition to the problematic of facility survey model creation from 3D point clouds with knowledge support. The suggested system is materialized via WiDOP project [*4*]. Furthermore, the created WiDOP platform is able to generate an indexed scene from unorganized 3D point clouds visualized within virtual reality modeling language [*5*].

The reminder of this paper is organized as follows: The next section describes briefly the most recent related works, followed by the prototype definition in section three. In section four, more focus on the domain ontology structure presenting the core behind WiDOP prototype will take place where we highlight the ontology structure and the created extension with the SWRL language to satisfy the target purpose. Section five presents a use cased materialized by the scene of the German rail way. Finally, we conclude and give insight on our future work in section six.

## II. BACKGROUND CONCEPT AND METHODOLOGY

The technical survey of facilities, as a long and costly process, aims at building a digital model based on geometric analysis since the modeling of a facility as a set of vectors is not sufficient in most cases. To resolve this problem a new standard was developed over ten years by the International Alliance for Interoperability (IAI). This standard, called IFC [*6*], considers the facility elements as objects that are defined by a 3D geometry and normalized semantic [*14*]. The problematic of 3D object detection and scene reconstruction including semantic knowledge was recently treated within different domain, basically the photogrammetry one [*7*], the construction one, the robotics [*8*] and recently the knowledge engineering one [*4*]. Modeling a survey, in which low-level point cloud or surface representation is transformed into a semantically rich model is done in three tasks where the first is the data collection, in which dense point measurements of the facility are collected using laser scans taken from key locations throughout the facility; Then data processing, in which the sets of point clouds from the collected scanners are processed. Finally, modeling the survey in which the low-level point cloud is transformed into a semantically rich model. This is done via modeling geometric knowledge, qualifying topological relations and finally assigning an object category to each geometry [*9*].

Concerning the geometry modeling, we remind here that the goal is to create simplified representations of facility components by fitting geometric primitives to the point cloud data [17]. The modeled components are labeled with an object category. Establishing relationships between components is important in a facility model and must also be established. In fact, relationships between objects in a facility model are useful in many scenarios. In addition, spatial relationships between objects provide contextual information to assist in object recognition [10]. Within the literature, three main strategies are described to rich such a model where the first one is based on human interaction with provided software's for point clouds classifications and annotations [11]. While the second strategy relies more on the automatic data processing without any human interaction by using different segmentation techniques for feature extraction [8]. Finally, new techniques presenting an improvement compared with the cited ones by integrating semantic networks to guide the reconstruction process [12].

*A. Manual survey model creation*

In current practice, the creation of facility model is largely a manual process performed by service providers who are contracted to scan and model a facility. In reality, a project may require several months to be achieved, depending on the complexity of the facility and the modeling requirements. Reverse engineering tools excel at geometric modeling of surfaces, but with lack of volumetric representations, while such design systems cannot handle the massive data sets from laser scanners. As a result, modelers often shuttle intermediate results back and forth between different software packages during the modeling process, giving rise to the possibility of information loss due to limitations of data exchange standards or errors in the implementation of the standards within the software tools [13]. Prior knowledge about component geometry, such as the diameter of a column, can be used to constrain the modeling process, or the characteristics of known components may be kept in a standard component library. Finally, the class of the detected geometry is determined by the modeler once the object created. In some cases, relationships between components are established either manually or in a semi-automated manner.

*B. Semi-Automatic and Automatic methods*

The manual process for constructing a survey model is time consuming, labour-intensive, tedious, subjective, and requires skilled workers. Even if modeling of individual geometric primitives can be fairly quick, modeling a facility may require thousands of primitives. The combined modeling time can be several months for an average sized facility. Since the same types of primitives must be modeled throughout a facility, the steps are highly repetitive and tedious [12]. The above mentioned observations and others illustrate the need for semi-automated and automated techniques for facility model creation. Ideally, a system could be developed that would take a point cloud of a facility as input and produce a fully annotated as-built model of the facility as output. The first step within the automatic process is the geometric modeling. It presents the process of constructing simplified representations of the 3D shape of survey components from point cloud data. In general, the shape representation is supported by CSG [15] or B-Rep [16] representation. The representation of geometric shapes has been studied extensively [15]. Once geometric elements are detected and stored via a specific presentation, the final task within a facility modeling task is the object recognition. It presents the process of labeling a set of data points or geometric primitives extracted from the data with a named object or object class. Whereas the modeling task would find a set of points to be a vertical plane, the recognition task would label that plane as being a wall, for instance. Often, the knowledge describing the shapes to be recognized is encoded in a set of descriptors that implicitly capture object shape. Research on recognition of facilities specific components related to a facility is still in its early stages. Methods in this category typically perform an initial shape-based segmentation of the scene, into planar regions, for example, and then use features derived from the segments to recognize objects. This approach is exemplified by Rusu et al. who use heuristics to detect walls, floors, ceilings, and cabinets in a kitchen environment [8]. A similar approach was proposed by Pu and Vosselman to model facility façades [18].

To reduce the search space of object recognition algorithms, the use of knowledge related to a specific facility can be a fundamental solution. For instance, Yue et al. overlay a design model of a facility with the as-built point cloud to guide the process of identifying which data points belong to specific objects and to detect differences between the as-built and as-designed conditions [19]. In such cases, object recognition problem is simplified to be a matching problem between the scene model entities and the data points. Another similar approach is presented in [20]. Other promising approaches have only been tested on limited and very simple examples, and it is equally difficult to predict how they would fare when faced with more complex and realistic data sets. For example, the semantic network methods for recognizing components using context work well for simple examples of hallways and barren, rectangular rooms [10], but how would they handle spaces with complex geometries and clutter.

*C. Discussion:*

The presented methods for survey modeling and object recognition rely on hand-coded knowledge about the domain. Concepts like "Signals are vertical" and "Signals intersect with the ground" are encoded within the algorithms either explicitly, through sets of rules, or implicitly, through the design of the algorithm. Such hard-coded, rule based approaches tend to be brittle and break down when tested in new and slightly different environments. Furthermore, it can be difficult to extend an algorithm with new rule or to modify the rules to work in new environments. Based on these observations, we predict that more standard and flexible representations of facility objects and more sophisticated guidance based algorithms for object detection instead of a standard one will open the way to significant improvement in facility modeling capability and generality.

### III. WiDOP PROTOTYPE

WiDOP platform is a Java platform presenting a knowledge based detection of objects in point clouds based on OWL ontology language, Semantic Web Rule Language, and 3D processing algorithms. It aims at combining geometrical analysis of 3D point clouds and specialist's knowledge to get a more reliable facility model. In fact, this combination allows the detection and the annotation of objects contained in point clouds. WiDOP prototype takes in consideration the adjustment of the old methods and, in the meantime, profit from the advantages of the emerging cutting edge technology. From the principal point of view, our system still retains the storing mechanism within the existent 3D processing algorithms, in addition, suggest a new field of detection and annotation, where we are getting a real-time support from the target scene knowledge. Add to that, we suggest a collaborative Java Platform based on semantic web technology (OWL, RDF, and SWRL) and knowledge engineering in order to handle the information provided from the knowledge base and the 3D packages results.

The field of the Deutsch Bahn railway scene is treated for object detection. The objective of the system consists in creating, from a set of point cloud files, from an ontology that contains knowledge about the DB railway objects and 3D processing algorithms, an automatic process that produces as output a set of tagged elements contained in the point clouds.

The process enriches and populates the ontology with new individuals and relationships between them. In order to graphically represent these objects within the scene point clouds, a VRML model file [*5*] is generated and visualized within the prototype where the color of objects in the VRML file represents its semantic definition. The resulting ontology contains enough knowledge to feed a GIS system, and to generate IFC file [*6*] for CAD software. As seen in Figure 1, the created system is composed of three parts.

- Generation of a set of geometries from a point could file based on the target object characteristics
- Computation of business rules with geometry, semantic and topological constrains in order to annotate the different detected geometries.
- Generation of a VRML model related to the scene within the detected and annotated elements

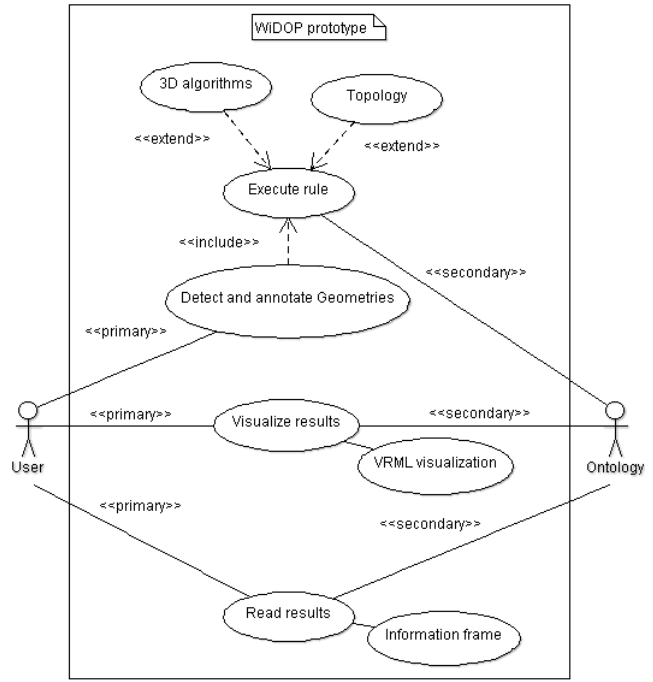

Figure 1. the WiDOP use case diagram

To rich such a target, three main steps aim at detecting and identifying objects are established:

- From 3D point clouds to geometric elements.
- From geometry to topological relations.
- From geometric and/or topological relations to semantic elements annotation.

As a first impression, the system responds to the target requirement since it would take a point cloud of a facility as input and produce a fully annotated as-built model of the facility as output. In the next, we focus on the core of the WiDOP prototype which is materialized via an ontology base structure to guide the 3D scene reconstruction process.

### IV. ONTOLOGY BASED PROTOTYPE

In recent years, formal ontology has been suggested as a solution to the problem of 3D objects reconstruction from 3D point clouds [*21*]. In this area, ontology structure was defined as a formal representation of knowledge by a set of concepts within a domain, and the relationships between those concepts. It is used to reason about the entities within that domain, and may be used to describe the domain. Conventionally, ontology presents a "formal, explicit specification of a shared conceptualization" [*22*].
Well-made ontology owns a number of positive aspects like the ability to define a precise vocabulary of terms, the ability to inherit and extends exiting ones, the ability to declare relationships between defined concepts and finally the ability to infer new relationship by reasoning on existing ones. Through the scientific community, the basic strength of formal ontology is their ability to reason in a logical way based on Description Logics DL. The last one presents a form of logic to reason on objects. In fact, despite the richness of OWL's set of relational properties, the axiom does not cover the full range of expressive possibilities for object relationships that we might find. For that, it is useful to declare a relationship in term of conditions or even rules. Some of the evolved languages are related to the semantic web rule language (SWRL) and advanced Jena rules [*23*]. SWRL is a proposal as a Semantic Web rules language, combining sublanguages of the OWL Web Ontology Language with the Rule Markup Language [*24*].

*A. Ontology schema*

This section discusses the different aspects related to the Deutsche Bahn scene ontology structure installed behind the WiDOP Deutsche Bahn prototype [*4*]. The domain ontology presents the core of WiDOP project and provides a knowledge base to the created application. The global schema of the modeled ontology structure offers a suitable framework to characterize the different Deutsche Bahn elements from the 3D processing point of view. The created ontology is used basically for two purposes:

- To guide the processing algorithm sequence creation based on the target object characteristics.
- To facilitate the semantic annotation of the different detected objects inside the target scene.

The created knowledge base related to the Deutsche Bahn scene has been inspired next to our discussion with the domain expert and next to our study based on the official Web site for the German rail way specification "http://stellwerke.de". The current ontology is divided onto three main parts: the Deutsche Bahn concepts, the algorithm concepts and the geometry concepts. However, they will be used with others to facilitate the object detection based on SWRL and the automatic annotation of Bounding Box geometry based on inference engine tools. At this level, no real interaction between human and the knowledge base is taken in consideration, since the 3D detection process algorithm and parameters are alimented directly from the knowledge base and then interpreted by the SWRL rules and Description Logics tools. The ontology is managed through different components of Description Logics. There are five main classes within other data and objects properties able to characterize the scene in question.

- Algorithm
- Geometry
- DomainConcept
- Characteristics
- Scene

The class DomainConcept can be considered the main class in this ontology as it is the class where the different elements within a 3D scene are defined. It was designed after the DB scene observation. It contains all kinds of elements, which have to be detected and is divided in two general classes, one for the Furniture and one for the Facility Element. However, the importance of other classes cannot be ignored. They are used to either describe the domain concept geometry and characteristics or to define the 3D processing algorithms within the target geometry. The subclasses of the Algorithm class represent the different developed algorithms. They are related to several properties which are able to detect. These properties (Geometric and semantic) are shared with the DomainConcept and the Geometry classes. By this way, a created sequence of algorithms can detect all the characteristics of an element while the Geometry class represents every kind of geometry, which can be detected in the point cloud scene.

The connection between the basic mentioned classes is carried out through object and data properties. There exist object properties for each mentioned activities. Besides, the object properties are also used to relate an object to other objects via topological relations. In general, there are five general object properties in the ontology which have their specialized properties for the specialized activities. They are

- hasTopologicRelation
- IsDeseignedFor
- hasGeometry
- hasCharacteristics,

Figure **2** demonstrates the general layout schema of the application.

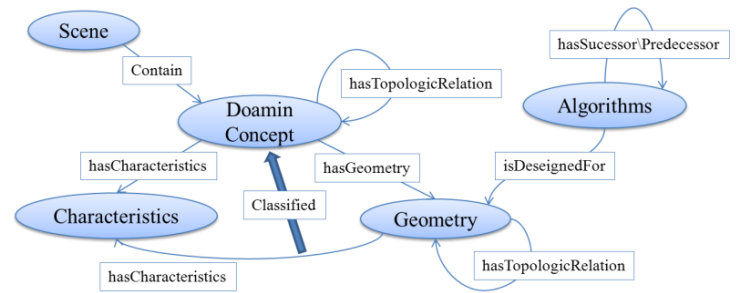

Figure 2. Ontology general schema overview

*B. Enrichment of the ontology within processing and topologic operations*

To support the defined use cases, two basic further layers to the semantic one are added to ontology in order to ensure the geometry detection and annotation process tasks. These operations are the 3D processing and topological relations qualification respectively.

*1) 3D processing operations*

The 3D processing layer contains all relevant aspects related to the 3D processing algorithms. Its integration into the WiDOP semantic framework is done by special Built-Ins.

They manage the interaction between the above mentioned layers and the semantic one. In addition, it contains the different algorithm definitions, properties, and the related geometries to the each defined algorithms. An important achievement is the detection and the identification of objects with specific characteristics such as a signal, indicator columns, and electric pole, etc. through utilizing their geometric properties. Since the information in point cloud data sometimes is unclear and insufficient.

*2) Topological operations*

The layer of the topological knowledge represents topological relationships between scene elements since the object properties are also used to link an object to others by a topological relation. For instance, a topological relation between a distant signal and a main one can be defined, as both have to be distant from one kilometer. The qualification of topological relations into the semantic framework is done by topological Built-Ins.

*C. Extension of SWRL with 3D processing and topological operations*

This section resumes the adopted approach to integrate the mentioned processing and topologic operation with help of the swrl language (Horn clauses) in order to define new knowledge (Classes and properties) related to the as built facility modeling. We recall that SWRL Built-ins allow further extensions within a defined taxonomy. In fact, it helps in the interoperation of SWRL with other formalisms by providing an extensible, modular built-ins infrastructure for Semantic Web Languages and knowledge based applications. For such a reason, we opt to be based on such a technology to extend the actual knowledge base within two basic Built-Ins: Topologic Built-Ins and Processing Built-Ins.

*1) Extension of standard SWRL with processing operations*

The first step aims at the geometric elements' detection. Thus, Semantic Web Rule Language within extended built-ins is used to execute a real 3D processing algorithm first, and to populate the provided knowledge within the ontology (e.g. Table 1). The "3D_swrlb_Processing: VerticalElementDetection" built-ins for example, aims at the detection of geometry with vertical orientation. The prototype of the designed Built-in is:

```
3D_swrlb_Processing:VerticalElementDetection (?Vert, ?Dir)
```

where the first parameter presents the target object class, and the last one presents the point clouds' directory defined within the created scene in the ontology structure. At this point, the detection process will result bounding boxes, representing a rough position and orientation of the detected object. Table 1 show the mapping between the 3D processing built-ins, which is computer and translated to predicate, and the corresponding class.

TABLE 1. 3D PROCESSING BUILT-INS MAPPING PROCESS

| 3D Processing Built-Ins | Correspondent Simple class |
|---|---|
| 3D_swrlb_Processing: VerticalElementDetection (?Vert,?Dir) | Vertical_BoundingBox(?x) |
| 3D_swrlb_Processing: HorizentalElementDetection (?Vert,?Dir) | Horizental_BoundingBox(?y) |

*2) Extension of standard SWRL with topologic operations*

Once geometries are detected, the second step, aims at verifying certain topology properties between detected geometries. Thus, 3D_Topologic built-ins have been added in order to extend the SWRL language. Topological rules are used to define constrains between different elements. After parsing the topological built-ins and its execution, the result is used to enrich the ontology with relationships between individuals that verify the rules. Similarly, to the 3D processing built-ins, our engine translates the rules with topological built-ins to standard rules, Table 2.

TABLE 2. EXAMPLE OF TOPOLOGICAL BUILT-INS

| Processing Built-Ins | Correspondent object property |
|---|---|
| 3D_swrlb_Topology:Upper(?x, ?y) | Upper(?x,?y) |
| 3D_swrlb_Topology:Intersect(?x, ?y) | Intersect (?x,?y) |

## V. CASE STUDY

For the demonstration of our prototype, 500 m from the scanned point clouds related to Deutsch Bahn scene in the city of Nürnberg was extracted. It contains a variety of the target objects. The whole scene has been scanned using a terrestrial laser scanner fixed within a train, resulting in a large point cloud representing the surfaces of the scene objects. Within the created prototype, different rules are processed, Figure 3. First, geometrical elements will be searched in the area of interest based on dynamic 3D processing algorithm sequence created based on semantic object properties, and then topological relations between detected geometries are qualified. Subsequently, further annotation may be relayed on aspects expressing facts to orientation or size of elements, which may be sufficient to finalize a decision upon the semantic of an object or on a fact expressing topological relationship or both of them.

This second step within our approach aims to identify existing topologies between the detected geometries. To do, useful topologies for geometry annotation are tested. Topological Built-Ins like `isConnected`, `touch`, `Perpendicular`, `isDistantfrom` are created. As a result, relations found between geometric elements are propagated into the ontology, serving as an improved knowledge base for further processing and decision steps. The last step consists in annotating the different geometries. Vertical elements of certain characteristics can be annotated directly. In more sophisticated cases, our prototype allows the combination of semantic information and topological ones that can deduce more robust results by minimizing the false acceptance rate. Finally, based on a list of SWRL rules, most of the detected geometries are annotated. In this example, among 13 elements are classified as Masts, 15 as Schaltanlage, three basic signals and finally, three secondary signals.

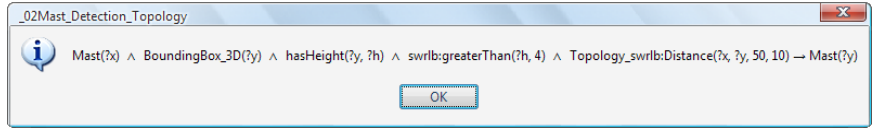

Figure 3. WiDOP prototype and example of used swrl rules within Built-Ins extention

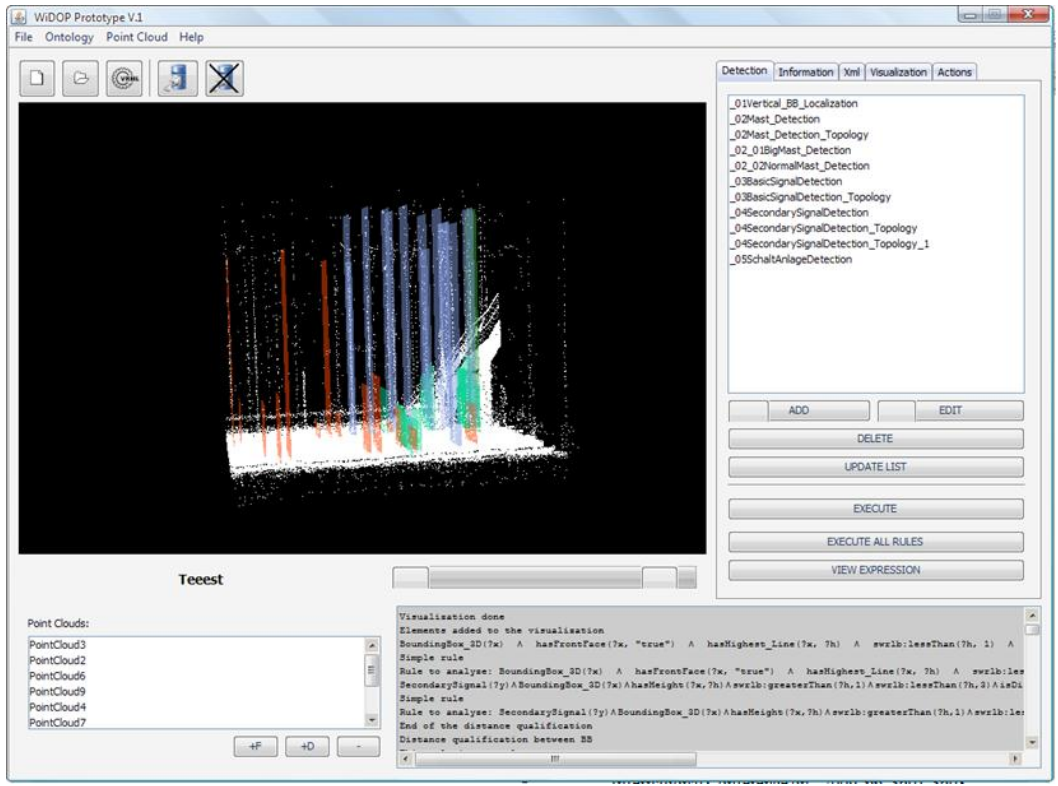

Figure 4. Detected and annotated elements visaliazation within VRML language

However, next to our experience, some limits are encountered. They are especially related very small elements detection and qualification where some noise on the ground still considered as semantic element. From our point of view, we think that the reason for such a false annotation is the lack of semantic characteristics related to such elements since until now; there is no real internal or external topology, neither internal geometric characteristic that discriminate such an element compared to others.

The created WiDOP platform offers the opportunity to materialize the annotation process by the generation and the visualization based on a VRML structure alimented from the knowledge base. It ensures an interactive visualization of the resulted annotation elements beginning from the initial state, to a set of intermediate states coming finally to an ending state, Figure 4 where the set of swrl rules are totally executed.

## VI. CONCLUSION AND FUTURE WORKS

We have presented an automatic system for survey information model creation based on semantic knowledge modeling. Our solution aims to perform the detection of objects from laser scanner technology by using available knowledge about a specific domain (DB). The designed prototype as simple, as efficient and intelligent it is since it takes 3D point clouds of a facility and produce fully annotated scene within a VRML model file. The suggested solution for this challenging problem has proven its efficiency through real tests within the Deutsche Bahn scene. The creation of processing and topological Built-Ins has presented a robust solution to resolve our problematic and to prove the ability of the semantic web language to intervene in any domain and create the difference.

Future work will include a more robust identification and annotation process of objects based on each object characteristics add to the integration of new 3D parameter knowledge's that can intervene within the detection and annotation process to make the process more flexible and intelligent.

ACKNOWLEDGMENT

This paper presents work performed in the framework of the research project funded by the German ministry of research and education under contract No. 1758X09. The authors cordially thank for this funding. Special thinks also for Hung Truong, Andreas Marbs, Ashish Karmacharya, Yoann Rous and Romain Tribotté for their contribution.